%Paper: hep-ph/9304230
%From: arzt@mich.physics.lsa.umich.edu (Christopher Arzt)
%Date: Thu, 8 Apr 1993 03:45:13 -0400

\input phyzzx

\def\theabstract{Effective Lagrangians, including those that are spontaneously
broken, contain redundant terms. It is shown that the classical equations of
motion may be used to simplify the effective Lagrangian, even when quantum
loops are to be considered.}

% DEFINITIONS
\font\bigboldiii=cmbx10 scaled\magstep3

\def\and{{\it\&}}
\def\half{{1 \over 2}}

\def\ocal{{\cal O}}
\def\dcal{{\cal D}}
\def\lcal{{\cal L}}
\def\lowti#1{_{{\rm #1 }}}
\def\lef{\lcal \lowti{ eff } }
\def\Dslash{\not\!\! D}
\def\lct{\lcal_{ct}}

\PHYSREV
\nopubblock{
\singlespace
\rightline{$\caps UM-TH-92-28$}
\rightline{hep-ph/9304230}}
{\titlepage
%\vskip -.3 in
\title{{\bigboldiii
Reduced Effective Lagrangians}}
\smallskip
\medskip
\medskip
\medskip
\titlestyle{{\twelvecp Christopher Arzt }}
{\twelvepoint\it\centerline{ Institute for Theoretical Physics, University of
California} \baselineskip=12pt \centerline{ Santa Barbara, CA 93106, USA}}
\baselineskip=12pt
\smallskip
\centerline{and}
\smallskip
{\twelvepoint\it\centerline{ Randall Laboratory of Physics, University of
Michigan} \baselineskip=12pt \centerline{ Ann Arbor, MI 48109-1120,
USA\foot{{\rm Present address}}} \baselineskip=12pt
\centerline{ARZT@mich.physics.lsa.umich.edu}
}
\smallskip
\smallskip
\smallskip
\singlespace
\abstract
\theabstract
\endpage}
\doublespace
%\sanseri

\chapter {{\caps Introduction}}

Within the framework of accelerator physics, there are two ways we may obtain
information on physics above the weak scale.  By building higher and
higher energy machines, we hope to directly observe new particles as the
energy of the machine passes the threshold of particle production.
At the SSC, for example, we hope
to find some direct evidence of a mass-generating mechanism or other
new phenomena.  We may also hope that high-precision measurements at lower
energies will provide indirect evidence of high-energy physics.
In these measurements we search for deviations
from the Standard Model; any such deviation is either evidence that
the Standard Model is incorrect or a signal of some new physics
on a higher energy scale.

If we accept the validity of the Standard Model, it becomes important to
ask how to characterize any possible deviations from it.  Although we do not
know the structure of physics beyond the Standard Model, we know that the
low-energy effects of the full higher-energy theory (valid at energies above
the mass  $\Lambda$ of some new particle) can be incorporated into an effective
Lagrangian
\REF\efflag{
H. Georgi and H. Politzer, {\sl Phys. Rev.} {\bf D14} (1976) 1829.
E. Witten, {\sl Nucl. Phys.} {\bf B104} (1976) 445.
E. Witten, {\sl Nucl. Phys.} {\bf B122} (1977) 109.
J. Collins, F. Wilczek, and A. Zee {\sl Phys. Rev.} {\bf D18} (1978) 242.
S. Weinberg, {\sl Phys. Rev. Lett.} {\bf 43} (1979) 1566.
F. Wilczek and A. Zee, {\sl Phys. Rev. Lett.} {\bf 43} (1979) 1571.
Y. Kazama and Y.P. Yao, {\sl Phys. Rev.} {\bf D21} (1980) 1116.
B. Ovrut and H. Schnitzer, {\sl Phys. Rev.} {\bf D21} (1980) 3369.
S. Weinberg, {\sl Phys. Lett.} {\bf 91B} (1980) 51.}
[\efflag].  For any given extension of the Standard Model we can write an
effective Lagrangian composed of only low-energy (Standard-Model) fields in  a
series of terms of higher and higher dimension:
$$
	\lef =	\sum_{n=-2}^{\infty} {1\over{{\Lambda}^n}}\thinspace
		\alpha_{\ocal}\thinspace \ocal^{(n+4)}
\eqn\eflag
$$
The operators $\ocal^{(n+4)}$ have dimension $[\hbox{mass}]^{(n+4)}$ and
contain only derivatives and fields with masses below $\Lambda$. If the
high-energy physics decouples
\REF\decou{
T. Appelquist and J. Carazzone, {\sl Phys. Rev.} {\bf D11} (1975) 2856.
K. Symanzik, {\sl Comm. Math. Phys.} {\bf34} (1973) 7.}
\ [\decou], then the sum starts at  $n=0$, and
$\ocal^{(4)}$ is the Standard Model.  All operators (not including those
related by the equations of motion) up to $\ocal^{(6)}$ have
been listed by Buchm\"uller and Wyler
\REF\bw{W. Buchm\"uller and D. Wyler, {\sl Nucl. Phys.} {\bf B268} (1986) 621;
{\sl Phys. Lett.} {\bf B197} (1987) 379.
See also:
C.N. Leung \etal\ {\sl Z. Phys.} {\bf C31} (1986) 433.
C.J.C. Burgess and H.J. Schnitzer, {\sl Nucl. Phys.} {\bf B228} (1983) 464.}
[\bw].
If the physics above scale $\Lambda$ does not decouple from the low energy
physics, then we can write our effective theory as a gauged chiral model
\REF\chirala{
T. Appelquist and C. Bernard, {\sl Phys. Rev.} {\bf D23} (1981) 425.
H. Georgi, {\sl Weak Interactions and Modern Particle Theory},
Benjamin / Cummings Publishing Co., Menlo Park, CA, USA (1984).}
\REF\chiralb{
A. Longhitano {\sl Nucl. Phys.} {\bf B188} (1981) 118.
S. Weinberg, {\sl Physica} {\bf A96} (1979) 327.}
\REF\gla{J. Gasser and H. Leutwyler, {\sl Ann. Phys.} {\bf 158} (1984) 142.}
\ [\chirala,\chiralb,\gla].  In
this case $\Lambda \sim 4\pi v$ [\chiralb], and there exist terms $\ocal^{(2)}$
of chiral  dimension 2. The (renormalization-scale-dependant) constants
$\alpha_{\ocal}$ determine the strength of the contribution of $\ocal$; they
are calculated by matching the Green's functions (or S-matrix elements) of the
effective Lagrangian to those of the full high-energy theory (for an
explanation see, for example,
\REF\onshell{H. Georgi, {\sl Nucl. Phys} {\bf B339} (1991) 339.}
\ [\onshell]).

The situation at hand is somewhat different.  We do not know the high-energy
theory, and so we cannot perform the matching to find the values for the
$\alpha$'s.  Whatever the high-energy theory is, though, it will generate
in the effective Lagrangian a
tower of terms $\ocal^{(n)}$ each obeying the symmetries of the theory.  So
we may parametrize all possible forms of new high-energy physics by
writing down an effective Lagrangian containing all operators
$\ocal$ which respect the symmetries of the theory.  Since higher
order terms are all suppressed by higher powers of ${ 1 \over \Lambda}$,
we can terminate the series at some point with negligible effects.

This still leaves a large number of terms to be considered.  The number can of
course be reduced by integration by parts; the action S is usually unchanged by
such a manipulation.  Further, the classical equations of motion can clearly be
used to remove terms when the effective Lagrangian is only to be used at tree
level
\REF\glb {J. Gasser and H. Leutwyler, {\sl Nucl. Phys.} {\bf B250} (1985) 93.}
\REF\derujula{A. De R\'ujula, M.B. Gavela, P. Hernandez, and E. Maas\'o,
CERN-TH.6272/91.} \ [\bw,\gla,\glb,\derujula].
Recently Georgi has shown that in certain cases the classical equations of
motion can be used even in the  quantum theory (in loop diagrams)  [\onshell].
Explicitly exempted from this in [\onshell] are  terms quadratic in the fields.
Also, gauge theories and spontaneously  broken theories are not fully
considered.  These limitations preclude the use of this simplification in the
Standard Model.  In this paper we will show that, quite generally, terms in an
effective Lagrangian that are connected by the equations of motion are
redundant.  They may be dropped from the effective Lagrangian without changing
observables.  A generalization of the equivalence theorem \REF\equivthm{J.S.R.
Chisholm, {\sl Nucl. Phys} {\bf 26} (1961) 469. S. Kamefuchi, L.
O'Raifeartaigh, and A. Salam, {\sl Nucl. Phys.} {\bf 28} (1961) 529. P.P.
Divakaran, {\sl Nucl. Phys.} {\bf 42} (1963) 235.  M.C. Berg\`ere and Y.M.P.
Lam, {\sl Phys. Rev. D} {\bf D13} (1976) 3247. H.S. Sharatchandra {\sl Ann.
Phys.} {\bf 116} (1978) 408. } \REF\ss{A. Salam and J. Strathdee, {\sl Phys.
Rev.} {\bf D2} (1970) 2869.} \REF\ren{R.E. Kallosh and I.V. Tyutin, {\sl Yad.
Fiz.} {\bf 17} (1973) 190 (reprinted as {\sl Sov. J. Nucl. Phys.} {\bf 17}
(1973) 98).} \ [\equivthm,\ss,\ren] shows that this is also true for quadratic
terms.  Spontaneously broken gauge theories present no new problems. The result
is that any effective Lagrangian can be brought into a canonical form
consisting of a reduced set of operators that are  gauge invariant and
unrelated by the classical equations of motion. We may choose such a set to
minimize the number of higher derivative terms; this is often (but not always -
for example see
\REF\gw{B. Grinstein and M. Wise, {\sl Phys. Lett.} {\bf B265} (1991) 326.}  \
[\derujula,\gw]) the most useful form.

\chapter{{\caps proof}}

The purpose of this section is to show that we may use the equations of motion
on any terms of dimension d $\geq 5$ in the effective Lagrangian.  In
particular, we show that any such term which contains $D^2 \phi$,  $D^\mu
F^I_{\mu\nu}$, or $\Dslash\psi$ gives contributions to the S matrix  identical
to those from a term with fewer derivatives, with which we may therefore
replace it.  (Here $\phi$ is a scalar, $F^I_{\mu\nu}$ is a gauge field
strength,  $\psi$ is a fermion, and $D_\mu$ is a covariant derivative.)  The
generalization will be clear.

Let $\lcal$ be an effective Lagrangian valid for energies below
$\Lambda$.  It can be
written as
$$
	\lef =	\sum_{n=-2}^{\infty} {1\over{{\Lambda}^n}}\thinspace
		\alpha_{\ocal}\thinspace \ocal^{(n+4)}
             \equiv  \sum_{n=0}^{\infty} \eta^n \lcal_n
\eqn\eq
$$
where $\eta$ is a small parameter (such as $1 / \Lambda$)
and $\lcal_0$ includes any terms with negative powers of $\eta$.
The operators $\ocal$ can all be chosen to be local; in the following
we assume this choice has been made.

As an example, consider $\lef = \half (\partial_\mu \phi)^2 - \half m^2 \phi^2
- \lambda \phi^4 + \eta g_1 \phi^6 + \eta g_2 \phi^3 \partial^2 \phi$.  By
making the shift of variables $\phi \rightarrow \phi + \eta g_2 \phi^3$ we
induce $\lef \rightarrow \half  (\partial_\mu \phi)^2 - \half m^2 \phi^2 -
\lambda' \phi^4 + \eta g_1' \phi^6 + \ocal(\eta^2)$. By the equivalence theorem
[\equivthm,\ss,\ren]\ the S matrix is unaffected by the change  in variables,
and so (to first order in $\eta$) we may choose to use the new  effective
Lagrangian in place of the original one.  In the following, we will  generalize
the equivalence theorem, showing that a similar procedure is  possible for any
effective Lagrangian.

In the following $\varphi_i$ stands for any of the various fields in the
theory. First consider an effective  Lagrangian containing a term like $\eta
T[\varphi] D^2 \phi$ where $T[\varphi]$ is any local function of any of  the
fields and their derivatives ($\eta$ is some appropriate power of $1 /
\Lambda$). Let $Z'[j_i]$ be the generating functional for the Green's functions
with the $j_i$ sources for each of the fields. Then (working for now only to
first order in $\eta$)
$$
Z'[j_i] = \int {\prod_l \dcal\varphi'_l\
       \exp\thinspace i\int{d^4x\left[\lcal_0' + (\lcal_1' - \eta T D^2 \phi')
		+ \eta T D^2 \phi' + \sum_i j_i \varphi'_i \right]}}
		+ \ocal(\eta^2).
\eqn\genfun
$$
The term in $\lcal_1'$ to be removed (that is, $\eta T D^2 \phi'$) has been
written explicitly.
We can now change variables so that $(\phi')^\dagger = \phi^\dagger + \eta T$.
(If the scalar is real, then we let $\phi' = \phi + \eta T$.)
This change is only a redefinition of a variable of integration, so we
expect $Z'$ (and therefore all Green's functions) to be unchanged.

Written in the shifted variable
$$\eqalign{
Z'[j_i] = \int \prod_l \dcal\varphi_l
	\left|{\delta(\phi')^\dagger \over \delta\phi^\dagger}\right|
       \exp\thinspace i\int d^4x\Bigg[\lcal_0' &+
	\eta T ({\delta\lcal_0' \over \delta\phi^\dagger} -
   	\partial_\mu {\delta\lcal_0' \over  \delta \partial_\mu\phi^\dagger})
	   + (\lcal_1' - \eta T D^2 \phi)  \cr
		&+ \eta T D^2 \phi
		+ \sum_{i} j_i \varphi_i
		+ j_{\phi^\dagger} \eta T \Bigg]
		+ \ocal(\eta^2).  \cr }
\eqn\genfunii
$$
In this equation we have expanded $\lcal\left(\phi^\dagger + \eta T\right)$ in
a  Taylor series about $\phi^\dagger$.  The shifted $Z'$ differs from \genfun\
in three ways:  there is a new Lagrangian, a Jacobian of the transformation,
and a new coupling to the source $j_{\phi^\dagger}$.

The change of the variable of integration we have performed in \genfun\
presumably has no effect. It is commonly known as the equivalence theorem
[\equivthm,\ren] that in many cases we may make the change of variables in {\it
only} the Lagrangian without changing the S matrix.  In other words, we may
remove the Jacobian of the transformation and the additional coupling to
$j_{\phi^\dagger}$ in \genfunii\ without changing the S matrix.  The only
effect on mass shell is to renormalize the Green's functions (though in
general the off-shell Green's functions are changed).   Statements of the
equivalence theorem for $\phi\rightarrow F(\phi)$ usually require $F$ to be a
point transformation, or $\phi\rightarrow \phi + F(\phi)$, where the expansion
of $F(\phi)$ begins with the term second order in $\phi$. For applications to
spontaneously broken theories, this would require the use of the shifted
field, destroying the (broken) symmetry.  These requirements
are in fact too demanding; we will see that the transformation $(\phi')^\dagger
= \phi^\dagger + \eta T$ leaves the S matrix unchanged for any function $T$ to
any order in $\eta$.  In the next three paragraphs we consider the three
differences between equations \genfun\ and \genfunii, respectively:
the change in the Lagrangian, the Jacobian of the transformation, and the new
coupling to the source $j_{\phi^\dagger}$.

The new Lagrangian is just the original Lagrangian plus $\eta T$ times the
classical equation of motion for $\phi^\dagger$.   The variable shift we have
performed respects the symmetries of the theory;   since $\phi^\dagger \phi$
and $T\phi$ are both invariant  under symmetry operations,  $\phi^\dagger +
\eta T$ transforms as $\phi^\dagger$ does. Because of this the new Lagrangian
explicitly retains all the symmetries of the original. If $\lcal_0$ has the
usual quadratic terms, then the new Lagrangian is $\lef + \eta T (-D^2\phi -
m^2 \phi + \hbox{terms with two or more fields)}$.  The first term cancels
$\eta T D^2\phi$ in $\lcal_1$.  Georgi [\onshell] has pointed out that since
the effective Lagrangian  contains {\it all} terms allowed by the symmetries of
the theory, each of remaining terms is of a form already present.  We can
absorb them by changing the coefficients $\alpha_\ocal$ of some terms already
present in $\lef$ (this is true to all orders in $\eta$).

Regardless of the the structure of $T[\varphi]$ (but assuming it is local) the
presence of the Jacobian has no effect on the theory.  It can be written as a
ghost coupling $\bar c c + \eta \bar c {\delta T \over \delta \phi} c$. We can
see that in any diagram containing ghosts, there will always be at least one
loop containing only ghost propagators. Assume without loss of generality that
$T$ has only one term.   \foot{If, on the contrary, $T = T_1 + T_2$, we  can
break the shift of variables into two parts:   $(\phi')^\dagger =
(\phi'')^\dagger + \eta T_1$, and then $(\phi'')^\dagger = \phi^\dagger + \eta
T_2$.  The net effect of the two  transformations is $(\phi')^\dagger =
\phi^\dagger + \eta T + \ocal(\eta^2)$.} The ghost Lagrangian from the shift
will have exactly two terms; one is  $\bar c c$, and the other will be a
kinetic term only if  $T = \partial^m \phi^\dagger$ for any m. So there can be
{\it either} a kinetic term for c (in which case the ghosts will not couple to
physical fields) {\it or} the ghosts will couple to  physical fields (in which
case there will be no kinetic term for the ghosts, which can therefore be
consistently disregarded when dimensional regularization is employed), but not
both.   Another way to see the point is to note that the effective theory is
valid only up to energies of order $\Lambda =  1 / \sqrt{\eta}$.  Let $T =
(\partial^2\phi + \lambda \phi^3)$, so the ghost Lagrangian is $\bar c( 1 +
\eta\partial^2 + 3\eta\lambda\phi^2) c$.  Now rescale $c\rightarrow c/
\sqrt{\eta}$. Even though the ghosts propagate, their mass is on the order of
the cutoff!  In any loop we can therefore expand the ghost propagators into the
numerator, and so loops consisting purely of ghosts will contribute only
quadratic (or more highly divergent) terms.  Any diagrams containing ghosts
will therefore not contribute to the S matrix. Note that for a more general
field transformation, the equivalence theorem does not hold because the ghosts
do not decouple [\ss]. For instance, for $\phi\rightarrow \partial^2\phi +
\lambda\phi^3$ the ghost Lagrangian  will be $\bar c \partial^2 c + 3 \lambda
\bar c c \phi^2$, which will have physical effects.  The transformation
necessary for the case at hand avoids this pitfall.

Again, whether or not $T$ is linear in $\phi$, the term $\eta
j_{\phi^\dagger}T$ has no effect on the S matrix.  Instead of
$$
{G^{(n)}}' = \bra{0} T\left[\phi(x_1)...\phi(x_n)\right] \ket{0}
$$
we have
$$
G^{(n)} =\bra{0} T\left[(\phi(x_1) + \eta T(x_1))...(\phi(x_n) + \eta
T(x_n))\right]\ket{0}.
$$
It can be seen diagrammatically that $G^{(n)} =
f^n(p) {G^{(n)}}' +$ (terms with fewer than n poles). The term $\eta
j_{\phi^\dagger} T$ has only the effect (on-shell) of multiplying each n-point
Green's function $G^{(n)}$ by $f^n(p)$, the nth power of some function of
momentum (see [\ren] for a full explanation in a  slightly more restricted
context).  Indeed, if $T$ is linear in $\phi$, even the off-shell Green's
functions are related to the original ones in this way.  In any case,
this multiplicative factor cancels out in the definition of the S matrix and
leaves all S-matrix elements unaffected.

The result is that $Z'$ gives the same S matrix as the generating functional
$$
Z[j_i] = \int {\prod_l \dcal\varphi_l\
       \exp\thinspace i\int{d^4x \left[\lcal_0
		+ (\lcal_1 - \eta T D^2 \phi)
		+ \sum_i j_i \varphi_i \right]}}
		+ \ocal(\eta^2).
\eqn\genfuniii
$$
The term $\eta T D^2 \phi$ has been removed,  the on-shell Green's functions
are the same up to a renormalization, and the (unknown) values of some
$\alpha_\ocal$'s have changed to linear combinations of the original
$\alpha_\ocal$'s (changing $\lcal'$ to $\lcal$).  This equivalence is true
regardless of the structure of the local function $T$.

The preceding comments are true also for a term $\eta T\Dslash\psi$. In this
case one makes the change of variables $\bar\psi \rightarrow \bar\psi +  \eta
T$.  The equations of motion contain $\Dslash\psi$ rather than
$D^2\phi$, and the rest follows similarly.

Finally, a term like $\eta T^\nu_a D^\mu F_{\mu\nu}^a$ is also redundant.
Here we must make the change of variables $A^\nu_a \rightarrow A^\nu_a + \eta
T^\nu_a$ (where A is any abelian or non-abelian Yang-Mills gauge field, $\mu$
and $\nu$ are Lorentz indices, and a,b, and c are symmetry-group indices). Note
that this change respects the local gauge symmetries; since the term $\eta
T^\nu_a D^\mu F_{\mu\nu}^a$ is gauge invariant, $T^\nu_a$ transforms like
$F^{\mu\nu}_a$.  Under a gauge transformation
$$\eqalign{
   A_a^\nu &\rightarrow A_a^\nu + \partial^\nu \Lambda_a
                              + gf_{abc}\Lambda_b A_c^\nu     \cr
   T_a^\nu &\rightarrow T_a^\nu + gf_{abc}\Lambda_b T_c^\nu   \cr}
\eqn\gt
$$
($f_{abc}$ are the structure constants of the symmetry group), so
$A^\nu_a + \eta T^\nu_a$ transforms just like $A_\nu^a$:
$$
   (A + \eta T)_a^\mu \rightarrow (A + \eta T)_a^\mu + \partial^\mu \Lambda_a
                              + gf_{abc}\Lambda_b (A + \eta T)_c^\mu.
\eqn\gt1
$$
The proof follows through unchanged from above, but now the action also
contains pieces whose job it is to fix the gauge.  The variable change
$A^\nu_a \rightarrow (A+\eta T)^\nu_a$ which
takes us from \genfun\ to \genfunii\ produces the following change in the
gauge-fixing term (using a simple choice as an example):
$$\eqalign{
\lcal_{GF} &= - {1 \over 2\xi} [f]^2 \cr
f =\partial_\mu A^\mu_a
  \ \ \ &\rightarrow \ \ \
f=\partial_\mu (A+\eta T)^\mu_a, \cr}
\eqn\gf
$$
and the following change in the Faddeev-Popov ghost term:
$$
\lcal_{FP} = \partial_\mu\omega_a^*(\partial^\mu\omega_a
                        + gf_{abc}\omega_b A^\mu_c)
\ \ \ \rightarrow \ \ \
\partial_\mu\omega_a^*(\partial^\mu\omega_a
                        + gf_{abc}\omega_b (A+\eta T)^\mu_c).
\eqn\fp
$$
The new FP ghost term is exactly that needed  $(\omega^*_a {\delta f \over
\delta \Lambda} \omega_a)$ for the new gauge-fixing term, and so the symmetry
is consistently fixed. After making the change of variables we can choose
instead to gauge-fix with the original gauge-fixing term, with the original FP
ghost term - the net effect is that our change of variables hasn't changed
$\lcal_{GF} + \lcal_{FP}$.

By repeatedly shifting variables, we can continue the process outlined above
for all redundant terms to all orders.  First we remove all order $\eta$ terms
containing the appropriate derivative forms.  Once we have removed
all possible derivative terms to order $\eta$,  we can continue the process
with derivative terms of order $\eta^2$; since the change of variables is $\phi
\rightarrow \phi +\eta^2 T$, all order $\eta$ terms will be unaffected.  In
this way we can successively remove derivative terms of the given form
order-by-order in $\eta$.

For a spontaneously broken theory the proof above carries through unchanged.
We may write the Lagrangian in terms of the shifted field $\phi$ (for
example
$\phi= { \phi_1 + v \choose \phi_2 }$) so that the (broken) symmetries
of the
theory are still apparent.  Then a term $\eta T[\varphi] D^2 \phi$ is
redundant, with the required shift in fields again being
$\phi^\dagger \rightarrow \phi^\dagger + \eta T$
(in the example this would mean $\phi_1 +v \rightarrow
\phi_1^* + v^* +  \eta T$ and $\phi_2^* \rightarrow \phi_2^* + \eta T$).
In this
case it is crucial that $T[\varphi]$ linear in the fields not be disallowed,
since upon expanding any shifted scalars $\phi$, $T[\varphi]$ may in
general
be a sum of terms, some of which are linear in the fields $\phi_1$ or
$\phi_2$.  That $T[\varphi]$
may be linear means all of these terms in $\eta T[\varphi] D^2 \phi$
are redundant, not just those with two or more fields.

\chapter{{\caps technical considerations}}

The preceeding manipulations are only formal, for two reasons.
Firstly, we have treated the path integral as if it were a simple integral
in the variable $\varphi$.  A rigorous treatment would discretize the
variables $\varphi(x)$ and write the path integral as an infinite product of
integrals over the discrete variables $\varphi_x$.  We would then  perform
the shift in variables, and
rewrite the result as a path integral over continuous variables.  Alternately,
we could change variables in the canonical operator formalism.  In this
way it is found
\REF\gj{J.L. Gervais and A. Jevicki, {\sl Nucl. Phys.} {\bf B110} (1976) 93.
S.F. Edwards and Y.V. Gulyaev, {\sl Proc. Roy. Soc.} {\bf A279} (1964) 229.}
\REF\shift{
H. Fukutaka and T. Kashiwa, {\sl Ann. Phys.} {\bf 185} (1988) 301.
J. Hietarinta, {\sl Phys. Rev.} {\bf D25} (1982) 2103.
W. Kerler, {\sl Nucl. Phys.} {\bf B129} (1978) 312.
W. Kerler, {\sl Lett. Nuov. Cim.} {\bf 23} (1978) 523.
R. Marnelius, {\sl Nucl. Phys.} {\bf B142} (1978) 108.}
\ [\gj,\shift] that, in general, the
formal manipulations above are actually incorrect; a discrepancy arises
at the two-loop level.  It will turn out, however, that this complication can
be dealt with.

In the operator formalism, the operator Hamiltonian $\hat H$
contains
non-commuting factors $\hat Q$ and $\hat P$.  It  is
related to the classical Hamiltonian $H$ by
$$
		\bra{q} \hat H(\hat Q,\hat P) \ket{p}_{QP}
	= { e^{ipq/\hbar} \over \sqrt{2 \pi \hbar}}  H(p,q)
\eqn\heqh
$$
The subscript QP indicates that H is ``QP ordered'', so that all
factors of $\hat Q$ are placed to the left of all factors of $\hat P$.  When
we make a change of variables, the new Hamiltonian
$ \hat H' = \hat H(f(\hat Q,\hat P),g(\hat Q,\hat P))$
is no longer QP ordered.  Because of the non-commuting factors,
$$
       \bra{q} \hat H \ket{p}_{QP} =
	\bra{q} \hat H' \ket{p} =
         \bra{q} (\hat H' + \hat H'_{new}) \ket{p}_{QP}
		 = { e^{ipq/\hbar} \over \sqrt{2 \pi \hbar}}
					( H' + H'_{new})
\eqn\eqn
$$
where the term after the first equals sign is not QP ordered, and $\hat
H'_{new}$ arises because $\hat P$ and $\hat Q$ do not commute.  We see that the
Hamiltonian in the new variables is not just $H'(f(Q,P),g(Q,P))$; there is an
additional term.

If, rather than the operator form, we consider a path integral over
classical fields, we see the same effect differently expressed.
Because of the stochastic nature of the path integral, terms
of order $\epsilon$ which we have naively neglected must be considered [\gj].
This is important, because the chain rule is not valid to order $\epsilon$;
this can be inferred from
${dq \over dt} = \lim_{\epsilon\rightarrow 0}
{ q(t+\epsilon) - q(t) \over \epsilon} =
\lim_{\epsilon\rightarrow 0}
[{dq \over dt} + \half\epsilon {d^2 q \over dt^2} +\ocal(\epsilon^2)]$.
When we change variables in the continuous Lagrangian, we ignore the
$\ocal(\epsilon)$ term.  But when we work carefully, with discrete variables,
these order $\epsilon$ terms in the integral are seen to contribute an
additional potential term in the limit $\epsilon \rightarrow 0$.
This is identical to $H'_{new}$.

In fact, this extra potential term is regulator dependant.  Salomonson
\REF\sal{P. Salomonson, {Nucl. Phys.} {\bf B121} (1977) 433} \ [\sal]\ noted
this by explicitly comparing Feynman graphs calculated before and after a
change of variables.  When using a cutoff, the extra potential was needed, but
when using dimensional regularization, no new term appeared. It can be shown
for a local field theory that the additional potential term is proportional to
$\delta^n(0)$, where $n+1$ is the number of space-time dimensions.  (In
operator language, this factor comes from the commutator of $\hat P$ and $\hat
Q$.  In the path integral version, it enters through additional volume elements
connected with the space integrals.)  The quantum mechanical path integral,
equivalent to a 0+1 dimensional field theory, has no such delta functions
factors. If the theory is regulated dimensionally, it would appear that these
terms leave the S matrix unchanged, and so the extra potential can be
disregarded.

Another possible source of error in the last section is the   manipulation of
divergent integrals. The conclusions reached above are strictly true only if we
assume the theory to have been regularized beforehand.  Let us  assume the
theory in \genfun\ has been rendered finite by dimensional regularization; as
$\epsilon$ approaches zero, the regulation is removed. In this case the
generating functionals in \genfun\ and \genfuniii\ give identical S-matrix
elements with identical $\epsilon$ dependence.  We can write the effective
Lagrangian  in \genfun\ as $\lef' + \lct'$ (a counter-term Lagrangian $\lct'$
has been extracted), so that the S-matrix elements from $Z'$ are UV finite
(\ie, they have no $\epsilon$ dependence).  We can then use the results of the
last section to show that all terms in $\lct'$ of dimension d $\geq 5$ which
contain $D^2 \phi$,  $D_\mu F^I_{\mu\nu}$, or $\Dslash\psi$ can be dropped in
favor of terms with fewer derivatives.  The reduced form of the full Lagrangian
will  contain no terms with these derivatives, either in $\lef$ or $\lct$, but
the S matrix will be the same as the that from $\lef'$ and $\lct'$, and so it
will be finite ($\epsilon$-independent).  The reduced set of effective
operators is therefore renormalizable, in the sense that counter terms of the
form already present in the reduced Lagrangian are sufficient to renormalize
all S-matrix elements.

In practice this type of renormalization may be cumbersome, as Green's
functions are divergent, and it is not until S-matrix elements are calculated
that the coefficients of the counter terms are apparent. In some cases it may
be easier to imagine employing all possible counter  terms, including those in
the form of terms not in the reduced $\lef$.   In this way we can clearly make
all Green's functions finite.  This approach has the disadvantage that terms
removed at one scale may reappear at another.  When running the couplings, some
that we removed will be reintroduced.  Of course, we may remove them again, at
the new scale, in the same way they were removed originally.  The relationships
among the $\alpha's$ will therefore be different at different scales, in just
such a way as to give the same $\alpha(\mu)'s$ in this method of
renormalization as in the other.

\chapter{{\caps Conclusions}}

Not all terms in an effective Lagrangian contribute independently to the S
matrix.  By using the classical equations of motion in the Lagrangian, the
number of terms which must be considered can be reduced while maintaining all
symmetries of the effective Lagrangian (including any broken symmetries). This
is true for all terms in an effective Lagrangian (including those quadratic in
fields) and for gauged and spontaneously broken theories.

A result of the discussion above is that, in the full effective Lagrangian,
some of the parameters $\alpha$ are redundant; it is only the the coefficients
of the terms in the maximally reduced effective Lagrangian that are completely
determined by the high-energy theory. For each equation of motion, there is one
arbitrary parameter, and we may exploit this ambiguity to choose values as we
see fit.  Additionally, it is clear that that the value of some $\alpha$ may
not be the same in two effective Lagrangians with differing sets of terms, and
so care must be taken when comparing estimates of an $\alpha$ to also compare
the forms of the Lagrangians used in calculations.

In practice this result is quite useful.  It is of great utility to be able to
work with as few terms in $\lef$ as  possible when calculating loop diagrams.
That the results apply in spontaneously broken theories, without destroying the
broken symmetry, makes them useful for calculating in the Standard Model
\REF\us{C. Arzt, M.B. Einhorn, and J. Wudka, University of Michigan preprint
UM-TH-92-17.} \ [\us], and these results have been assumed in, for example,
[\derujula]. Generally, it is easiest to calculate with an effective Lagrangian
in which the number of derivatives is minimized, though other choices are
possible.  Indeed other choices of operators are sometimes more useful, for
example when the derivative terms are more tightly constrained by experiment
than are those with which we would replace them \ [\derujula], or when the
derivative terms in the effective Lagrangian are expected, based on some
knowledge of the high-energy theory, to be small [\gw].

\chapter{{\caps Acknowledgements}}
I am very grateful to M.B. Einhorn for many valuable discussions, helpful
suggestions, and critical readings of the manuscript.  I also thank J. Wudka
for useful discussions of the subject.
Portions of this work were supported by the Department of Energy and by the
National Science Foundation (Grant No. PHY89-0435).

\medskip
\medskip
\medskip
\medskip
\refout
\bye